\documentclass[preprint,aps,showpacs,preprintnumbers,amsmath,amssymb,superscriptaddress,floatfix]{revtex4}

\usepackage{graphicx}
\usepackage{color}
\usepackage{dcolumn}
\usepackage{bm}

\def\bit{\begin{itemize}}
\def\eit{\end{itemize}}

\def\beq{\begin{equation}}
\def\eeq{\end{equation}}
\def\bea{\begin{eqnarray}}
\def\eea{\end{eqnarray}}

\begin{document}

\title{
\textbf {An experimental program for demonstrating precision jet energy measurement
at the ILC}
}

\author{Rajendran~Raja}
\email{raja@fnal.gov}
\affiliation{Fermi National Accelerator Laboratory, P. O. Box 500, Batavia, IL 60510}

\date{\today}

\begin{abstract} 
We outline a physics program at Fermilab that would significantly
improve our ability to understand the behavior of hadronic showers in
calorimeters. This would involve a two-pronged approach designed to
measure particle production cross sections of hadronic beams on
several nuclei to improve shower simulation programs and the
validation of the improved shower simulation predictions using test
beams (including tagged neutral beams) in which calorimeter modules
built using various technologies are deployed in the test beam. Such a
program would be of immediate benefit to the efforts to design and
build optimal calorimeters for the International Linear Collider.
\end{abstract}

\thispagestyle{empty}
\maketitle



\section{Introduction}

Hadronic shower simulation programs depend on theoretical models of
the strong interaction for generating events that form part of the
shower cascade. All these events occur at low enough momentum scales
that ensure that they cannot be calculated from first principles using
the QCD Lagrangian which is currently only useful in the perturbative
regime. The result is that we have a series of event generation models
with varying theoretical assumptions that are employed in shower
programs that have their own internal parameters which must be tuned
to describe particle production data on thin target nuclei to fix
their internal parameters.

The data on which the tuning is done are mostly obtained from single
particle spectrometers and form a set of varying quality spanning a
time period over of thirty years. Most models have sufficient
flexibility that their parameters can be adjusted to describe the
single variable distributions of the data on which they are tuned.

In ILC calorimeters, both the longitudinal and transverse shapes of
showers produced by charged and neutral particles need to be
understood. If we can describe the single particle transverse and
longitudinal calorimeter energy depositions properly and understand
them, we can hope to design the calorimeters that have the appropriate
resolution to satisfy the needs of the ILC physics groups.

When we employ models to predict hadronic showers in calorimeters,
however, we have to repeatedly generate events and propagate the
particles in the media that make up the calorimeter. Correlations
(momentum-multiplicity correlations, rapidity correlations,
longitudinal-transverse momentum correlations) between individual
particles in an event become important as well as the overall
inclusive single variable spectra used for tuning the models. The
various models will thus diverge in their predictions of showers since
the effects of the systematics of the models will be magnified during
their repeated use during the showering process.

It is impossible to tune the shower programs using newly acquired
calorimeter data, since it is difficult to correlate calorimeter data
with the fine structure of the models which generate events at the
hadron-nucleus level. What is done currently by some collider
experiments is to tune the parameters of the models to test beam data
taken on their already designed calorimeters and to use the tuned
models as interpolating devices to simulate events in their full
calorimeter. Such tuned models will only work as interpolators in
their limited area of tuning but will fail when extrapolated to other
materials or momentum regimes or experiments.

The demands of ILC calorimetry are much more stringent than this. One
needs to use the models to predict the behavior of showers in
calorimetry yet to be designed and then to optimize such calorimetry
to obtain hitherto unachieved jet energy resolutions. In order to do
this, one needs to improve the status of the simulators by obtaining
new multi-particle production data at the particle-nucleus level with
excellent statistics, particle identification and acceptance. With
such data, the systematics of the models can be controlled to such an
extent that the calorimeter shower predictions can be put on a more
robust basis.

While this program of improving the shower simulators is going on, we
propose to have a program of test beam activity that will test various
calorimeter schemes. One needs to decide whether the Particle Flow
Algorithm (PFA)~\cite{pfa} is viable or not. This requires the use of tagged
neutral beams as well as charged particle test beams using which the
behavior of hadronic showers is studied in detail in the
calorimeters. The role of compensated calorimetry also should be
investigated much more thoroughly, since compensation allows for
better linearity and resolution in the calorimeter. Linearity of
response is important for both PFA and other algorithms. As we will
show subsequently, it may be possible to design calorimeter schemes
which provide high resolution spatial information cheaply that can be
used for both the PFA as well as compensated calorimetry.  We propose
that Fermilab embark on a detector R\&D scheme to investigate such
schemes.

\section{Improving Hadronic Shower Simulators}
\subsection{Hadronic Production Models and systematics}

The current state of hadronic shower simulation programs was examined
in great detail at the Hadronic Shower Simulation Workshop~\cite{hssw6}
held at Fermilab in September 2006. As mentioned, all the simulator
programs must rely on models of non-perturbative QCD to make
predictions, since we have not yet succeeding in calculating
non-perturbative processes from first principles using the QCD
lagrangian. These models have free parameters which are tuned to
existing data.

\begin{figure}[tbh!]
\centerline{\includegraphics[width=\textwidth]{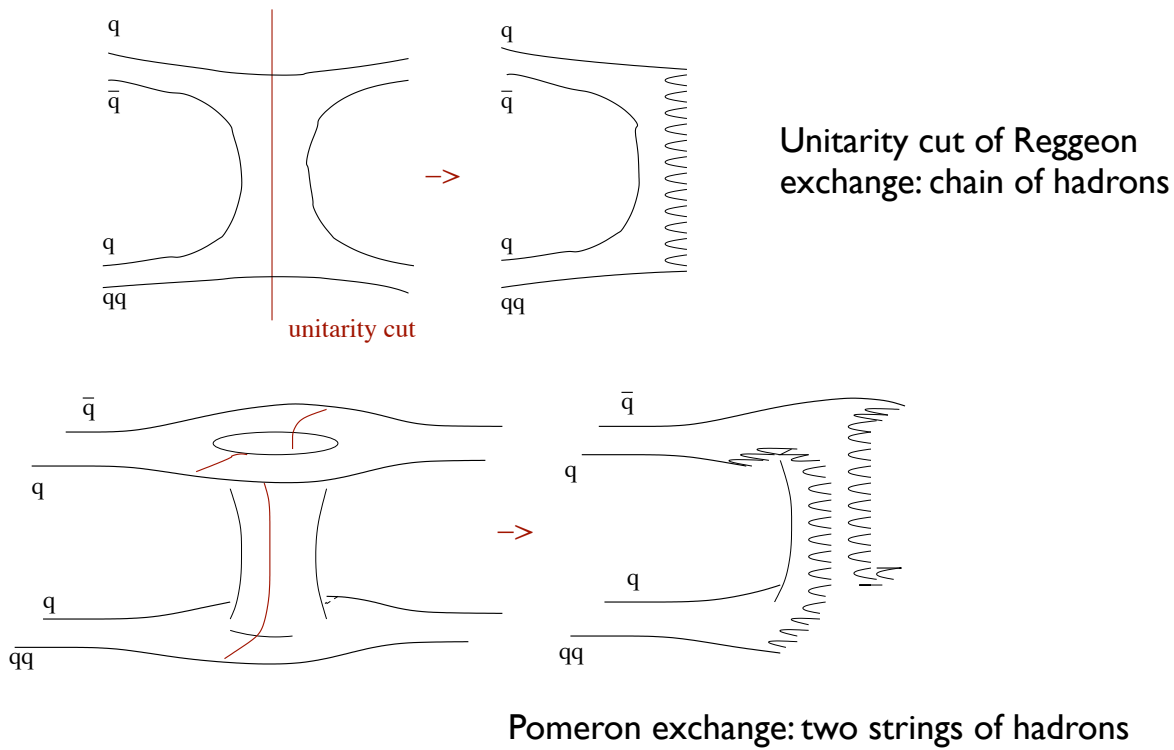}}
\caption
{The topological quark diagram for meson scattering and pomeron
scattering. The elastic scattering diagram is cut to produce two
strings for a single pomeron exchange. This approach is the basis of
the Dual Parton Model employed in DPMJET and QGSM.~\cite{engel}}
\label{pomeron}
\end{figure}

In order to illustrate the model dependence, let us consider in some
detail the event generator programs DPMJET (Dual Parton Model
JET)~\cite{dpmjet} and QGSM (Quark Gluon String
Model)~\cite{qgsm}. They both utilize the dual parton model of
hadronic production~\cite{capella} which utilizes the pomeron and
reggeon exchanges to describe soft processes. These processes are
illustrated in Figure~\ref{pomeron}. The reggeon exchange diagram can
be used to describe a sum of s-channel resonances as well as a sum of
t-channel exchanges as these are dual to each other topologically. The
pomeron diagram is topologically different from the reggeon exchange
planar diagram. When we employ the optical theorem, we can cut the
elastic scattering amplitude to reveal the parton structure of the
exchange to describe the total cross section. Thus, the reggeon
exchange contribution to the total cross section is described by a
single quark gluon string where as the pomeron exchange contribution
is described by two such strings.

\begin{figure}[tbh!]
\centerline{\includegraphics[width=\textwidth]{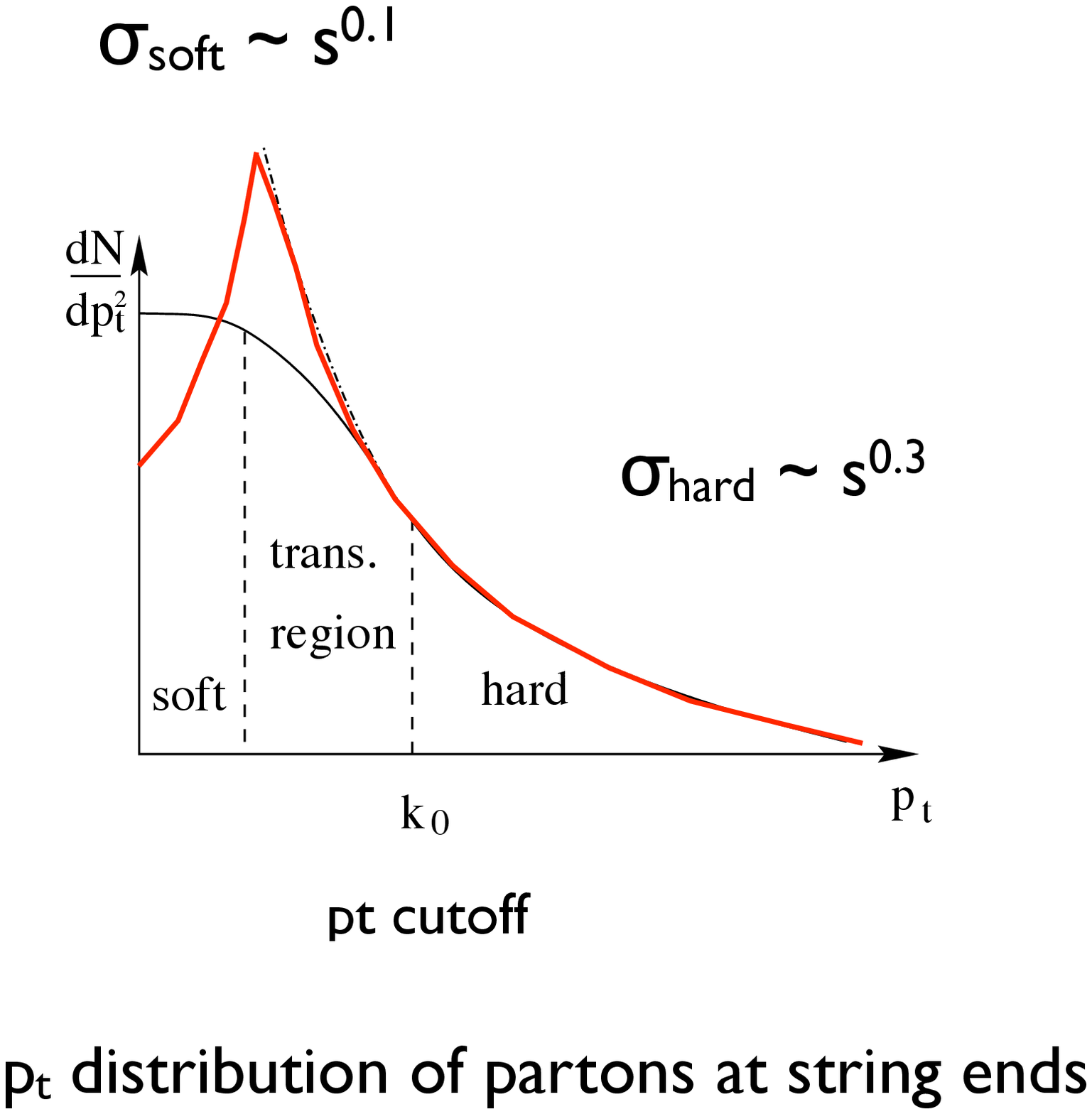}}
\caption
{Matching the energy dependence of the soft processes and the hard
processes is done arbitrarily and very carefully~\cite{engel}.}
\label{pomeron_matching}
\end{figure}

At high $Q^2$, these strings are described by the (perturbative)
parton models and structure functions, whereas at low $Q^2$, the
reggeon/pomeron approach is employed. The boundary between these two
formalisms is somewhat arbitrary and the joining of the two approaches
in itself contains free parameters as illustrated in
Figure~\ref{pomeron_matching}.
Finally, none of these models conserve unitarity per se, and unitarity
is imposed in DPMJET and QGSM  
by using the eikonal approximation. As we go to higher
center of mass energy (LHC energies), the number of pomerons exchanged
has to be increased to fit the increase in multiplicity.
Parton fragmentation into hadrons is an area of considerable
uncertainty and is a source of a great deal of systematic differences
between the models. Many different hadronization schemes are used each
containing a number of tunable parameters.
Nuclear fragmentation is also handled differently by different
programs. As we will show later, the fragmentation of the nucleus and
neutron emission as well as nuclear binding energies have to be
simulated well in order to understand the behavior of
calorimeters. Table~\ref{models} shows the various simulation 
programs~\cite{sim}
represented at the Hadronic Shower Simulation Workshop and the models
employed by them. It is clear that there is a plethora of models to
describe particle production and that none of these models will describe
the data perfectly, since there does not currently exist a complete
theory of non-perturbative QCD.

\begin{table}[tbh]
\caption[]{
Hadronic Shower Simulation Programs represented at the Hadronic Shower
Simulation Workshop and the models employed by them.
\label{models} 
}
\begin{center}
\begin{tabular}{|c|l|l|}
\hline
Program & Event Generator Models & Nuclear Break up models\\
\hline
Fluka05 & Isobar model (below few GeV)  & PEANUT (Includes GINC) \\
        & own version of DPM + hadronization & Generalized InterNuclear Cascade \\
\hline
Geant4 & QGS + Fritiof String model $> 20 GeV$ & Geant4 Pre-compound model \\
       & Bertini Cascade Model $< 10 GeV$  & Bertini evaporation model \\
       & Binary Cascade model  & Chiral Invariant Phase Space model (CHIPS) \\
 &  Low Energy Parametrized Models \& & $< 20 MeV$ Nuclear break-up libraries\\
       & High Energy Parametrized Models (GHEISHA origin) & \\
\hline
MARS15 & Inclusive event generator & Generalized intra-nuclear cascade\\
       & CEM03, LAQGSM03 Quark-Gluon String model & evaporation and fission models \\
\hline
PHITS & Jet AA Microscopic Transport Model (JAM) $> 20 MeV$ & Neutrons done as in MCNP \\
      & Jaeri Quantum Molecular Dynamics model JQMD & JQMD \\
\hline
MCNPX & Fluka79 or LAQGSM   & Intra Nuclear Cascade models \\
      &                     & Bertini, ISABEL, CEM, INCL4..\\
\hline
\hline
\end{tabular}
\end{center}
\end{table}
\subsection{The case for obtaining more particle production data}
The particle production data on which these models are tuned has
been obtained by numerous experiments over the last 30 years.  A
large number of these experiments are single arm spectrometers and
suffer from large systematic errors introduced by the fact a) it is
hard to calculate the acceptance of a single arm spectrometer since
the acceptance depends on the vertex position (hard to determine if
only a single final state particle is measured) and also the cross
section one is trying to measure. and b) Corrections due to decay of
$K^0_S$ and $\Lambda$ particles are hard to make.  The cross sections
measured by single arm spectrometers are in addition discrete in
transverse momentum for the various angular settings of the single
arm. Most importantly final particle correlations and multiplicity
distributions are not measured.

\begin{figure}[tbh!]
\centerline{\includegraphics[width=\textwidth]{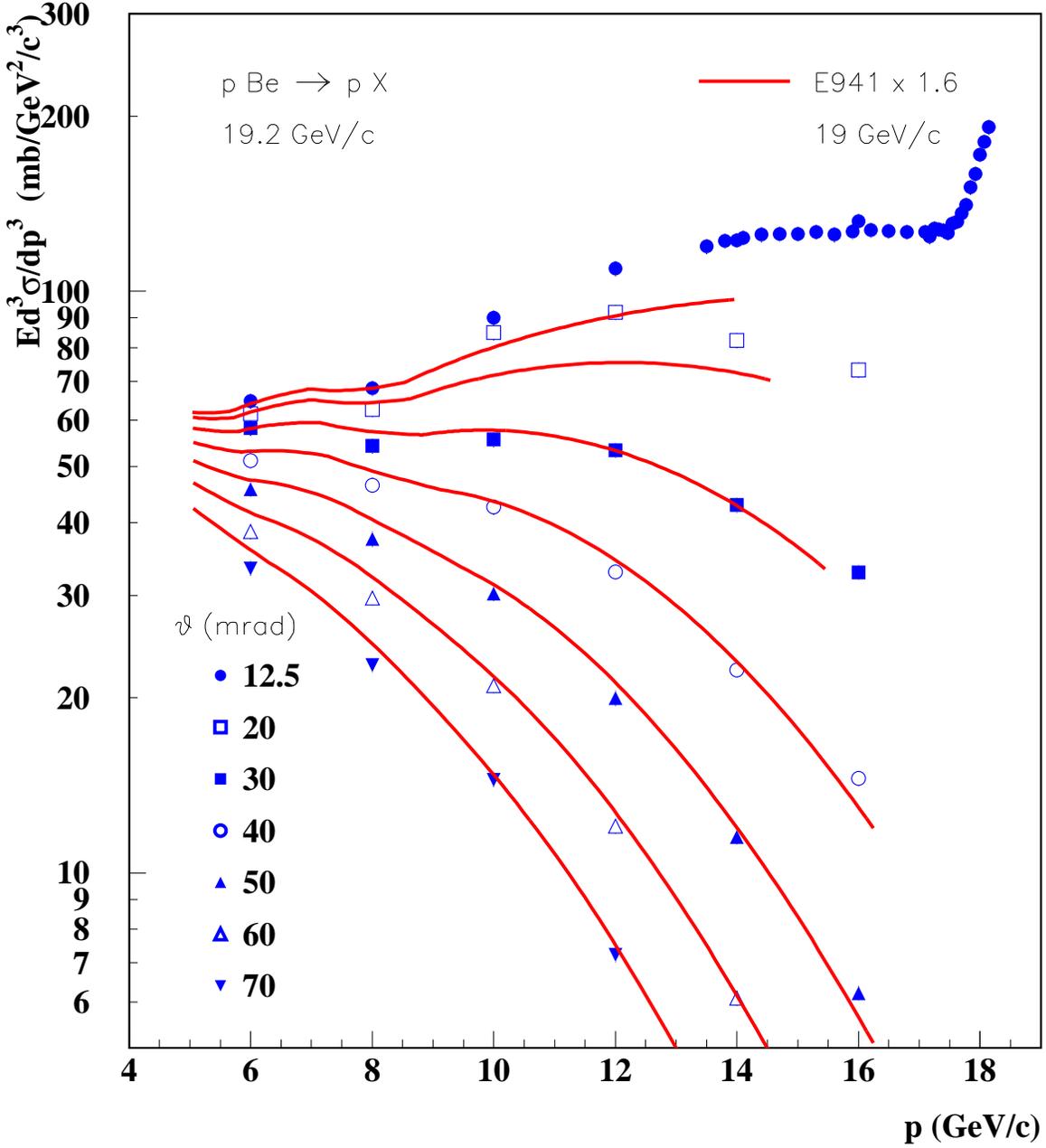}}
\caption
{Comparison of 19 GeV/c proton beryllium data between two different
experiments. There is a normalization difference of 1.6 between the
two.}
\label{allaby-bear-be}
\end{figure}

\begin{figure}[tbh!]
\centerline{\includegraphics[width=\textwidth]{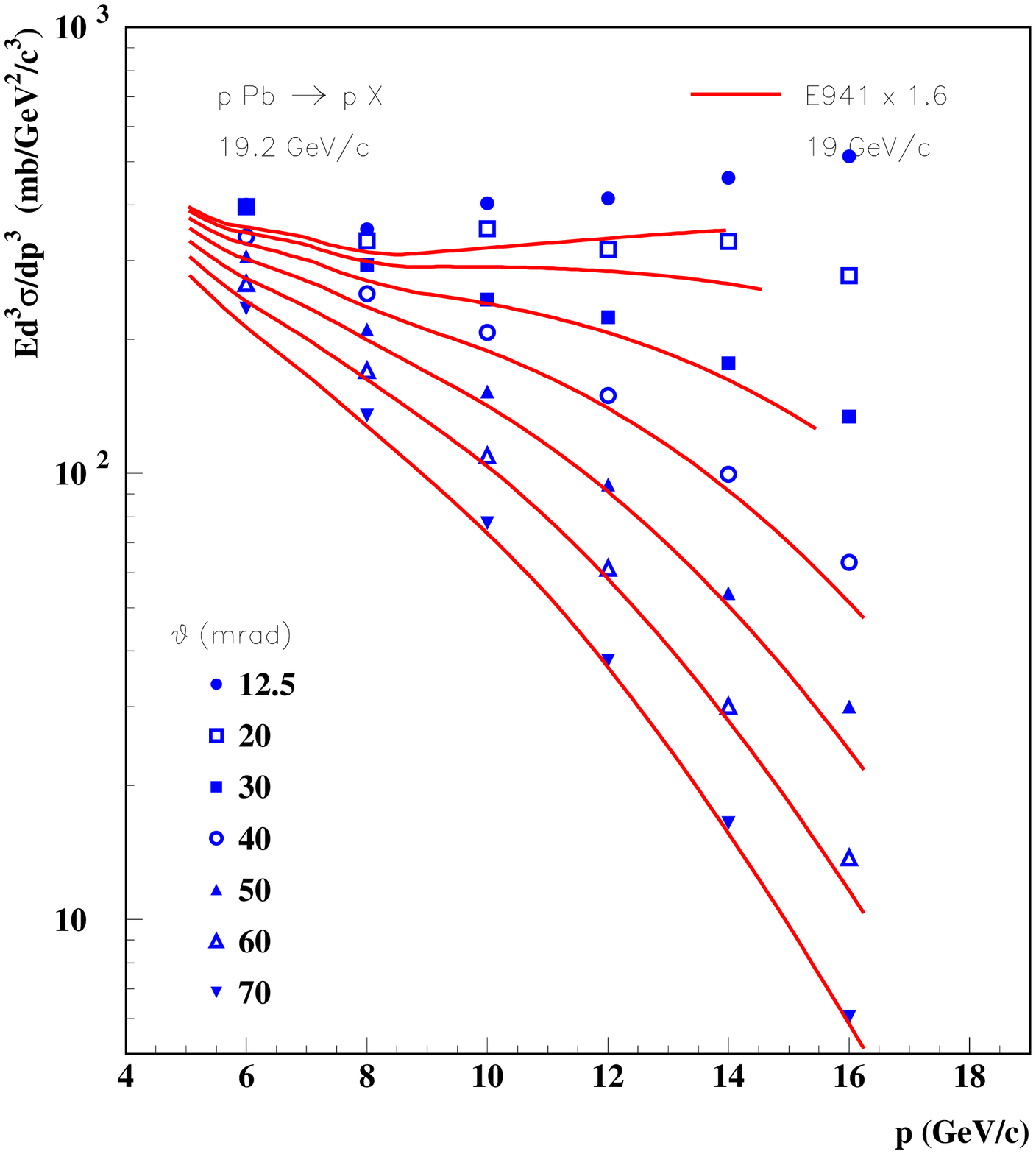}}
\caption
{Comparison of 19 GeV/c proton lead data between two different
experiments. There is a normalization difference of 1.6 between the
two.}
\label{allaby-bear-pb}
\end{figure}

Figure~\ref{allaby-bear-be} compares invariant inclusive cross section
for two experiments (Allaby et al (1970)~\cite{allaby}  
and BNL-E941  (2001)~\cite{e941}
for the reaction p+Be$\rightarrow$p+X  for 19~GeV/c incident 
protons as a function of the angle of the final state protons. E941 data has to be 
normalized by a factor 1.6 to agree with Allaby et.al.
Figure~\ref{allaby-bear-pb} compares invariant inclusive cross section
for the same two experiments 
for the reaction p+pb$\rightarrow$p+X  for 19~GeV/c incident 
protons as a function of the angle of the final state protons. E941 data has again 
to be  normalized by a factor 1.6 to agree with Allaby et.al.
All the models have been tuned to available data and describe the
distributions reasonably well. The only proviso is that the
distributions are usually functions of one variable only $x_F$ or
$p_T$ and do not really attempt to describe the particle
correlations. These correlations are of great importance in describe
a shower accurately, since a shower with $n-$ generations will magnify
the systematic errors $n$ fold. With a view to testing various
programs against each other, a set of benchmarks were set before the
Hadronic Shower Simulation Workshop~\cite{striganov}. We  detail here some of
the problem areas.

\begin{figure}[tbh!]
\centerline{\includegraphics[width=\textwidth]{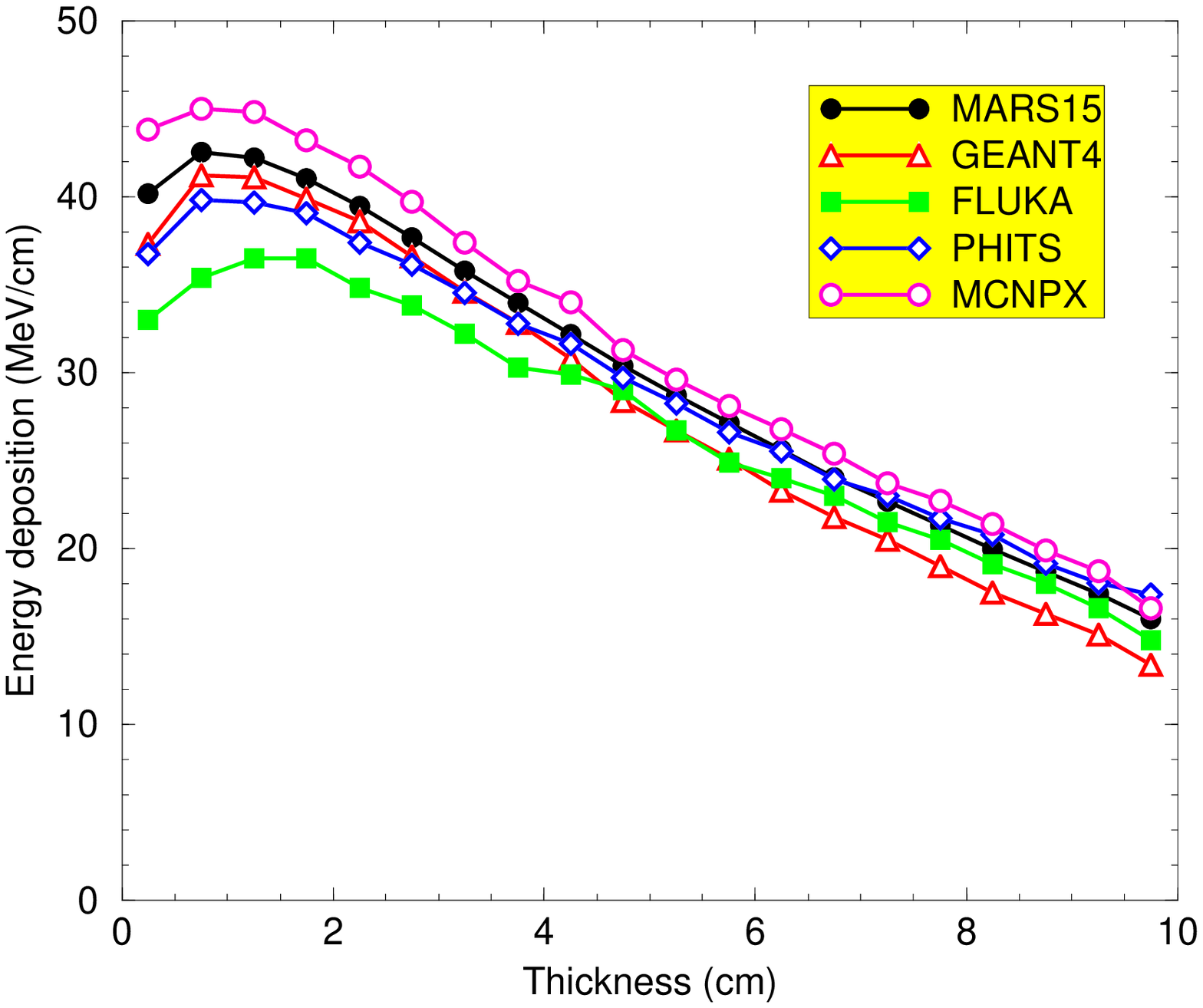}}
\caption
{Energy deposit profile as a function
of longitudinal depth of 1~GeV/c protons on a 10 cm Tungsten rod of 1
cm radius. Plotted are comparisons between MARS, GEANT4,PHITS and
MCNPX.}
\label{bench-f10a}
\end{figure}

\begin{figure}[tbh!]
\centerline{\includegraphics[width=\textwidth]{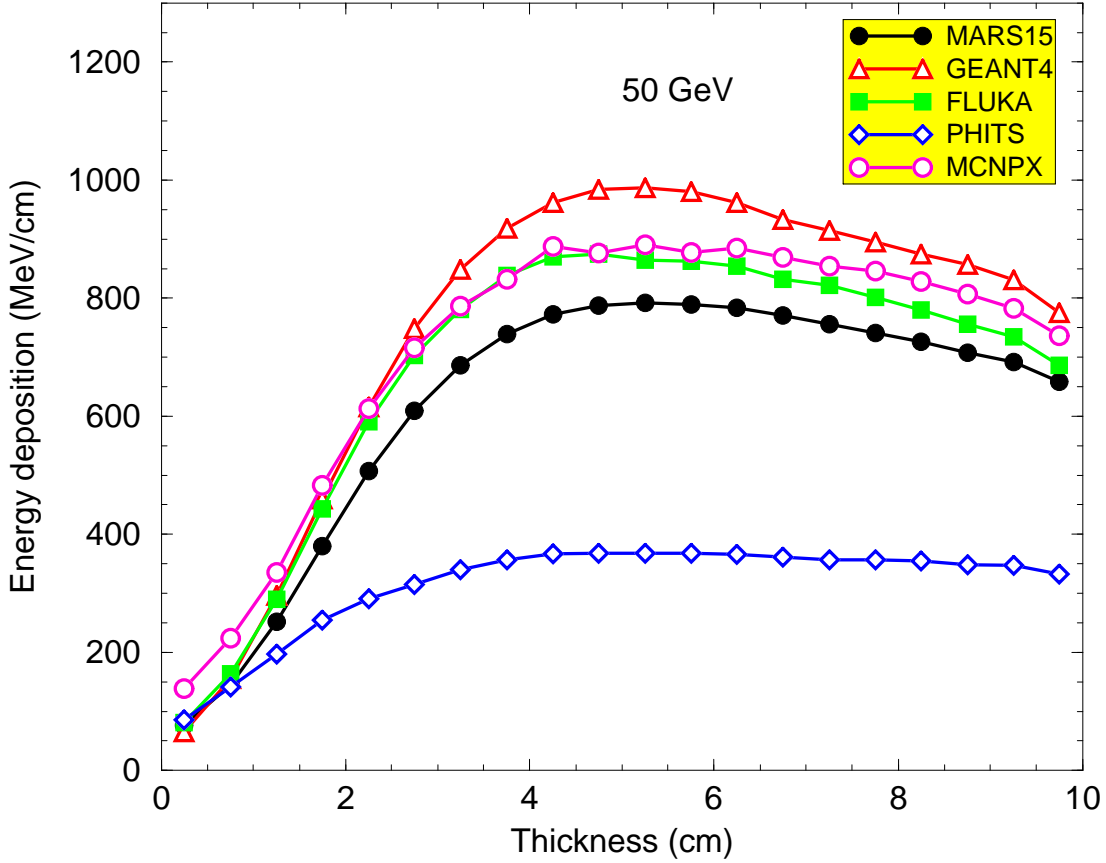}}
\caption
{Energy deposit profile as a function
of longitudinal depth of 50~GeV/c protons on a 10 cm Tungsten rod of 1
cm radius. Plotted are comparisons between MARS, GEANT4,PHITS and
MCNPX.}
\label{bench-f10b}
\end{figure}

Figure~\ref{bench-f10a} shows a comparison of the energy deposit as a
function of longitudinal depth in a 10~cm Tungsten rod of radius 1~cm
as predicted by the programs MARS15, GEANT4, PHITS and MCNPX for
1~GeV/c incident protons. There is significant disagreement between
the programs. 
Figure~\ref{bench-f10b} shows the same plot but with  50~GeV/c incident
protons. 
The discrepancies between the predictions between the various models 
are still significant.

\begin{figure}[tbh!]
\centerline{\includegraphics[width=\textwidth]{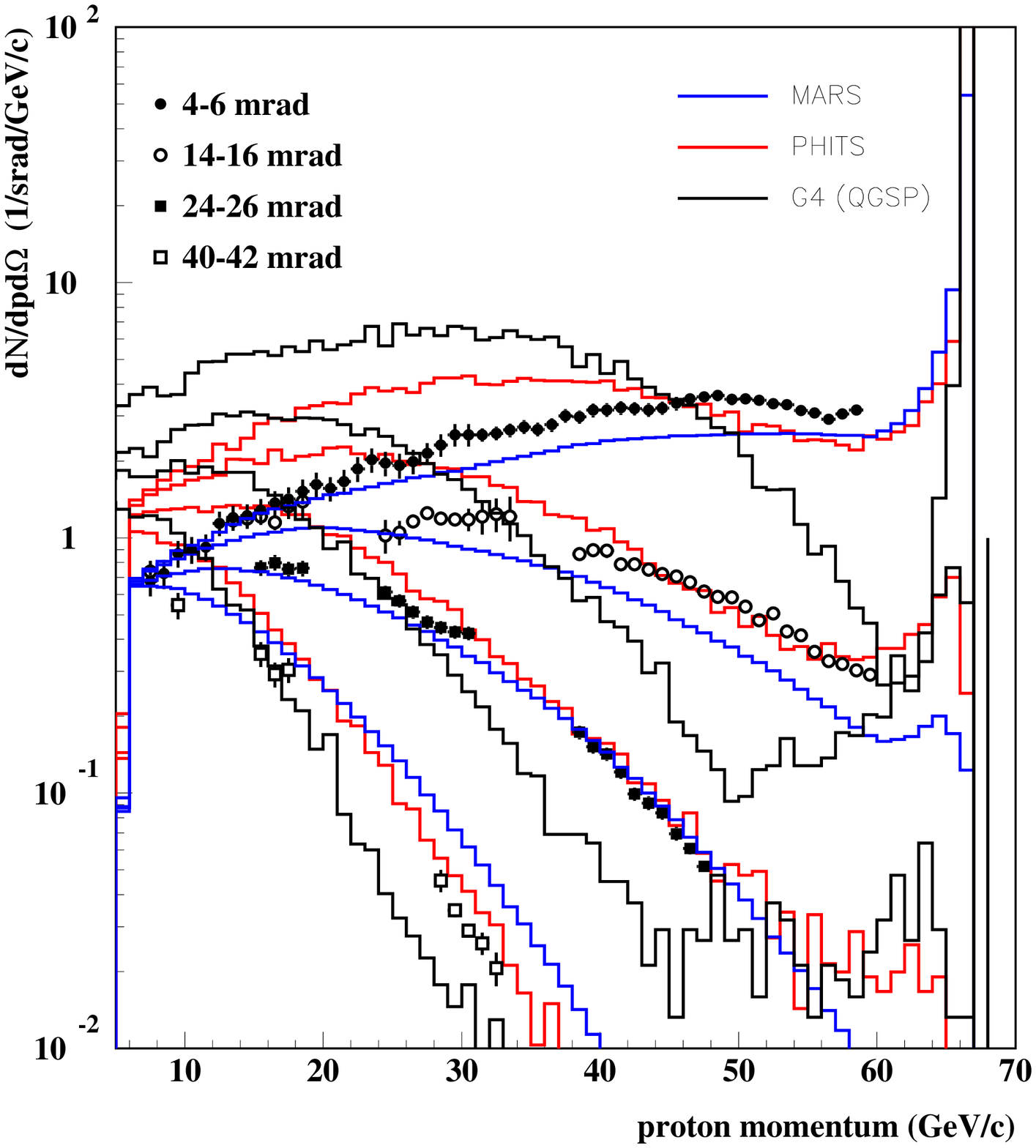}}
\caption
{Comparison of proton production on a thick aluminum target using a
beam of 67~GeV/c protons as a function of final state production
angle. Plotted are data (symbols) and predictions of MARS, PHITS, and
GEANT4 (using generator QGSP) (histograms). There is disagreement
between models themselves and models to data.}
\label{prot3}
\end{figure}

\begin{figure}[tbh!]
\centerline{\includegraphics[width=\textwidth]{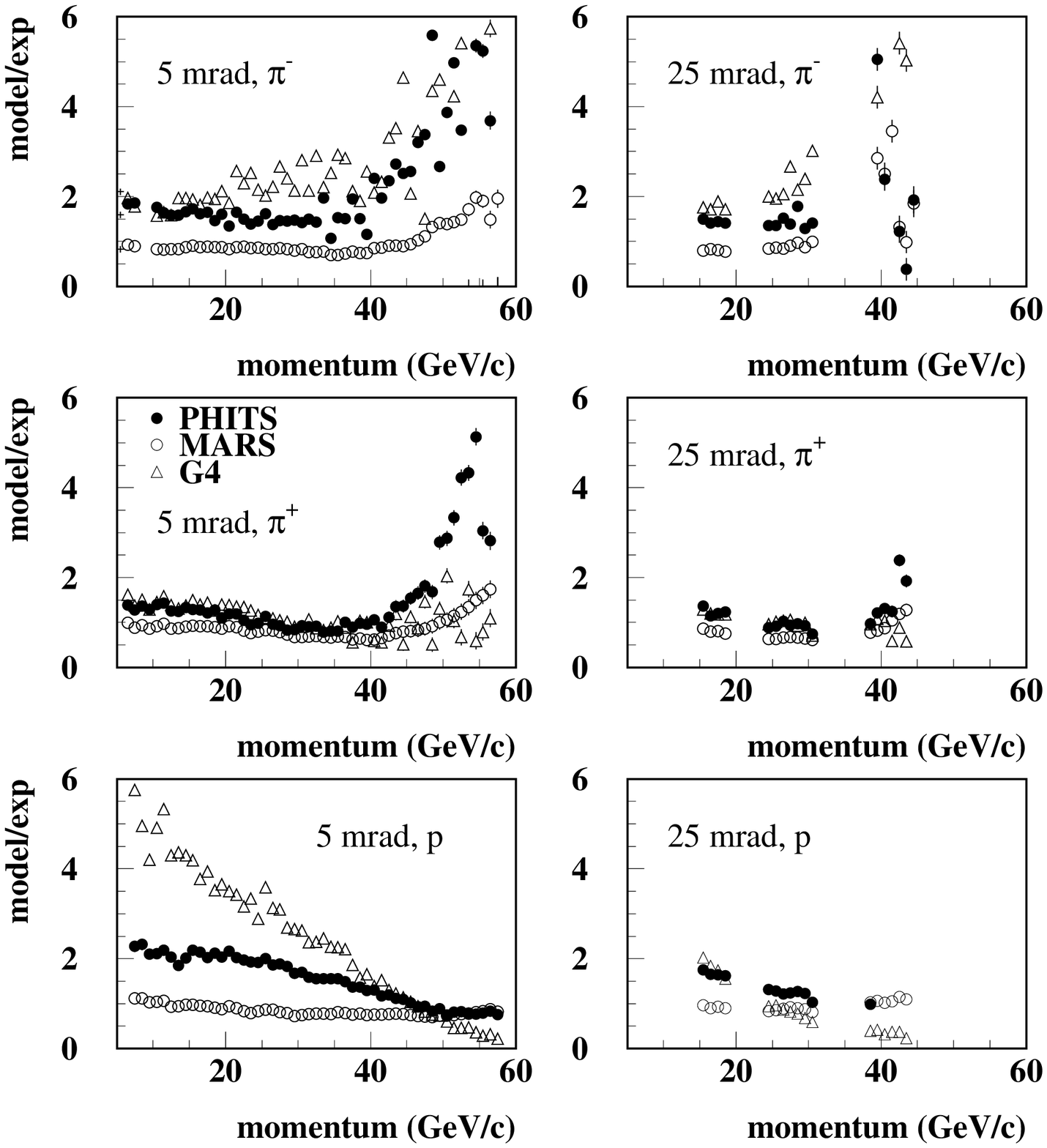}}
\caption
{Comparison of $\pi^\pm$ and proton production on a thick aluminum
target using a beam of 67~GeV/c protons as a function of final state
production angle. Plotted are the ratios of predictions of MARS,
PHITS, and GEANT4 (using generator QGSP) to data . Open triangles
represent Geant4/data, closed circles PHITS/data and open circles
MARS/data. Ratios of 5-6 means there is 500-600\% disagreement between
the specific model and data at that point of phase space!}
\label{rr}
\end{figure}

Figure~\ref{prot3} shows the results of another benchmark test where
the program predictions for inclusive proton production are compared
to data available for 67~GeV/c incident protons on a thick aluminum
target. The predictions, plotted as a function of the angle of the
final state proton disagree with each other and the data. Similar
plots are available for final state pions and kaons with similar
degree of discrepancies. We note that this test of the simulators is a
crucial benchmark for predicting neutrino spectra for experiments such
as MINOS, NOVA and MINERVA.
Figure~\ref{rr} shows the ratio of data/generators (PHITS, MARS and
GEANT4) for final state particles $\pi^+, \pi^-$ and $p$ for the
angles 5 and 25 mrad as a function of outgoing particle momentum.
Discrepancies of the order of a factor of 5 or 6 are evident in some parts of phase space.
It must be evident now that ability of the models to predict particle
production in conditions where thick targets are used is severely
under question. It was the unanimous opinion of the simulator experts
who participated in the Hadronic Shower Simulator Workshop that high
quality multi-particle data with good statistics and particle
identification is needed to improve simulators further.

\begin{figure}[tbh!]
\centerline{\includegraphics[width=\textwidth]{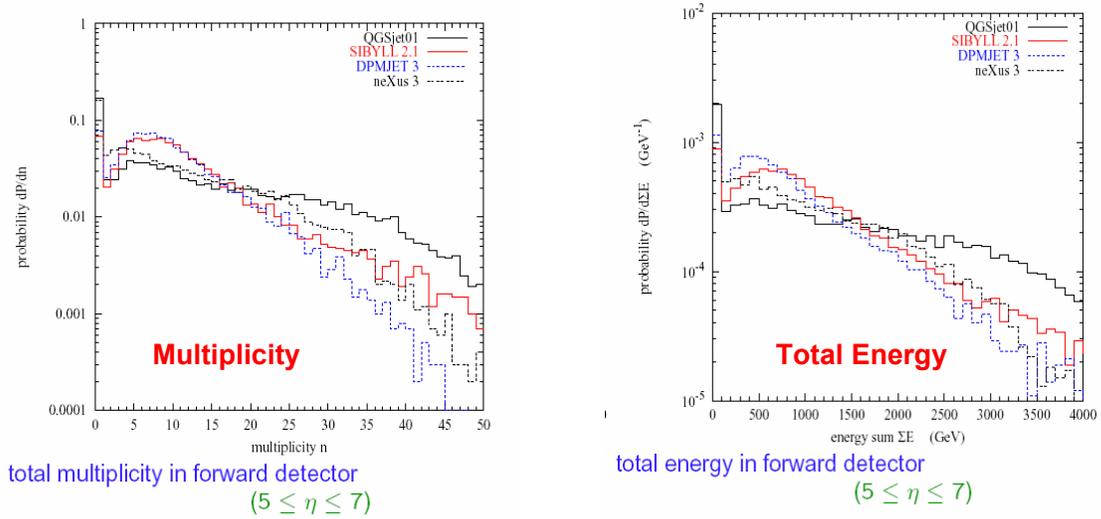}}
\caption
{Comparison of multiplicity and energy distributions in a forward
detector at the LHC $5<\eta<7$ as predicted by various
simulators. Plot courtesy of the TOTEM experiment~\cite{lami}}
\label{lami}
\end{figure}

It is not only when the target thickness is changed further that the
simulators run into trouble.  Similar situation holds when one enters
unchartered domains in center of mass energy. Figure~\ref{lami} shows
the prediction of multiplicities and energy distribution of particles
in a slice in pseudo-rapidity of $5<\eta<7$ at the LHC of various
simulators including QGSM, SYBILL and DPMJET. This plot is obtained
from the TOTEM experiment at the LHC. there is significant
disagreement in the shapes of the distributions between the simulator
programs.

\subsection{Possibilities for obtaining high quality data}

The MIPP experiment~\cite{mipp}
 has acquired high quality particle production data
on the nuclei H, Be, C, Al, Bi and U with 6 beam species ($\pi^\pm,
K^\pm$ and $p^\pm$) at various beam momenta ranging from 20 GeV/c to
120 GeV/c. MIPP has acquired 18 million events in total on these
targets. MIPP utilizes a TPC which accepts all the particles going
forward in the laboratory. Final state particles are identified
through all of phase space using a combination of $dE/dx$, time of
flight, multi-cell Cerenkov and RICH technologies. The events are 
currently being analyzed and papers are expected this year.

The rate of data acquisition in MIPP is limited to 20~Hz by the
electronics of the TPC which is 1990's vintage. With more modern
electronics, it is possible to upgrade the apparatus to run at 3000~Hz
in a cost effective manner. The details of the proposal can be found
on the MIPP website~\cite{mipp-upgr}. With this increase in speed, it
would be possible to acquire 5 million events per day, without any
increase in the number of protons being expended by the Main
Injector. These numbers are based on a 4 second Main Injector spill
every 2 minutes with a 42\% assumed downtime in beam delivery.  MIPP
Upgrade proposes to obtain 5 million events per nucleus on the nuclei
$H_2, D_2, Li, Be, B, C, N_2, O_2, Mg, Al, Si, P, S, Ar, K,Ca, Fe,
Ni,Cu, Zn, Nb, Ag, Sn, W, Pt,$ \\
$Au, Hg, Pb, Bi,U, Na, Ti, V, Cr, Mn, Mo,I, Cd, Cs$ and $Ba$. 
Each nucleus requires a single day of running. The
data obtained will be put on random access event libraries that are
indexed by beam species, beam energy, charged multiplicity and missing
neutral mass. These event libraries contain charged particle
information and events will contain missing neutrals. The random
access library which will fit on 36~Gigabytes of disk can be used
directly by the simulation program to look up events close to the
particles interacting in a shower. The looked up event can be scaled
to the appropriate interacting particle momentum. Missing neutrals can
be simulated by using sub-events that have no missing neutrals and
converting measured charged particles to neutral particles in a way
that conserves isospin. Such an approach will reduce the model
dependence of simulation programs and can be rapidly implemented 
once the DST's are obtained. The feasibility of the random access
library scheme can be evaluated even before MIPP upgrade data is
available by using the existing simulators to produce simulated thin
target data and using this simulated data to populate the library.  The
test would be to determine how well the library reproduces the
simulated data. Such work can be undertaken by people in the
simulation groups. Expertise in random access databases 
would be very valuable here.

A parallel approach will be to fit
the models to the new data, but since there is no exact theory, there are
no exact models and it may be difficult to fit all this data with a
single set of model parameters.

It should be pointed out that improving hadronic shower simulators
will benefit a broader community than the ILC effort. The neutrino
program, the collider programs, cosmic ray air showers and atmospheric
neutrino communities have systematics associated with shower
simulations that need to be brought under better control.

\section{Calorimeter Resolution}
The resolution of hadronic calorimeters has been extensively
studied~\cite{wigmans} and the effects of unequal responses to
hadronic and electromagnetic energy has been extensively reported. A
pion of 1~GeV energy interacting in lead on average deposits only
0.478~GeV as ionization and loses 0.414~GeV as binding energy involved
in nuclear break-up, 0.126~GeV into evaporation neutrons and
0.032~GeV into target recoil energy. Electromagnetic showers deposit
all their energy as ionization. This leads to an $e/h$ ratio that is
significantly larger than unity unless steps are taken to compensate
for the missing energy in hadronic reactions. This imbalance in
response is the source of two problems in hadron calorimetry namely a
degradation in energy resolution and non-linearity of response. The
degradation in energy resolution is the result of fluctuations in the
EM/Hadron energy from shower to shower due to the random nature of the
showering process. 

While reconstructing the shower, under the
assumption that one cannot tell apart the EM energy from the hadronic
energy, one is forced to use a single set of calibration 
constants for each shower
which results in a degradation of energy resolution. 

\subsection{ Constant term in resolution due to non-compensation}

The unequal response to electromagnetic and hadronic energies results 
in a constant term in the resolution function as can be seen trivially by
\begin{equation}
 E_{vis} = E_{em} + f E_{had}
\end{equation}
$E_{vis}$ 
is the energy deposited 
as $dE/dx$ in a single hadron shower in the calorimeter,
$E_{em}$ is the energy deposited as electromagnetic energy in the shower
and $E_{had}$ is the energy deposited as hadronic energy in the shower, 
and  $f$ is the $e/h$ ratio. Let $E_{live}$ be the energy observed in the live sampling layers of the calorimeter. Then
\begin{equation}
 E_{live} = E_{vis}/\mu
\end{equation}
where $\mu$ is the inverse sampling fraction of the calorimeter. Let
$E_{true}$ be the true energy of the parent hadron and let $\lambda$ be
the fraction of the energy in hadronic energy and $1-\lambda$ be the
fraction in $EM$ energy. $\lambda$ fluctuates from event to event. Then
one can write for a single event
\begin{equation}
 E_{live} = E_{vis}/\mu = \frac{(1-\lambda)E_{true} + f\lambda E_{true}}{\mu}
\end{equation}
This can be re-written
\begin{equation}
 E_{live} = \frac{((1-\lambda(1-f))E_{true}}{\mu}
\end{equation}
$E_{true}/\mu$ can be thought of as the ideal energy observable in a 
calorimeter with $e/\pi=1.0$. This then yields
\begin{equation}
 E_{live} = ((1-\lambda(1-f))E_{live}^{ideal}
\end{equation}
Taking logarithms and differentiating to calculate variances, one gets,
\begin{equation}
\frac {\sigma_{ E_{live}}}{E_{live}} = 
\frac{\sigma_{E_{ideal}}}{E_{ideal}} 
\oplus \sigma_{\lambda} <\frac{(1-f)}{1-\lambda(1-f)}>
\end{equation}
The term $\frac{\sigma_{E_{ideal}}}{E_{ideal}} $ will scale as
$1/\sqrt(E)$ where as the term $\sigma_{\lambda}
<\frac{(1-f)}{1-\lambda(1-f)}>$ will act as a constant term that vanishes 
in a perfectly compensating calorimeter with $f=1$. 
The brackets $<>$ denote r.m.s over events, and the 
symbol $\oplus$ denotes addition in quadrature. For high energies,
the constant term dominates the resolution and should be made as small
as possible while designing the calorimeter.

\subsection{Non-linearity in energy response due to non-compensation}

The non-linearity
results from the fact that the fraction of EM energy that hadronic
showers deposit increases with the energy of the parent hadron,
because EM showers do not produce further hadrons where as hadrons can
continue to give more EM energy further in the cascade.

\begin{figure}[tbh!]
\centerline{\includegraphics[width=\textwidth]{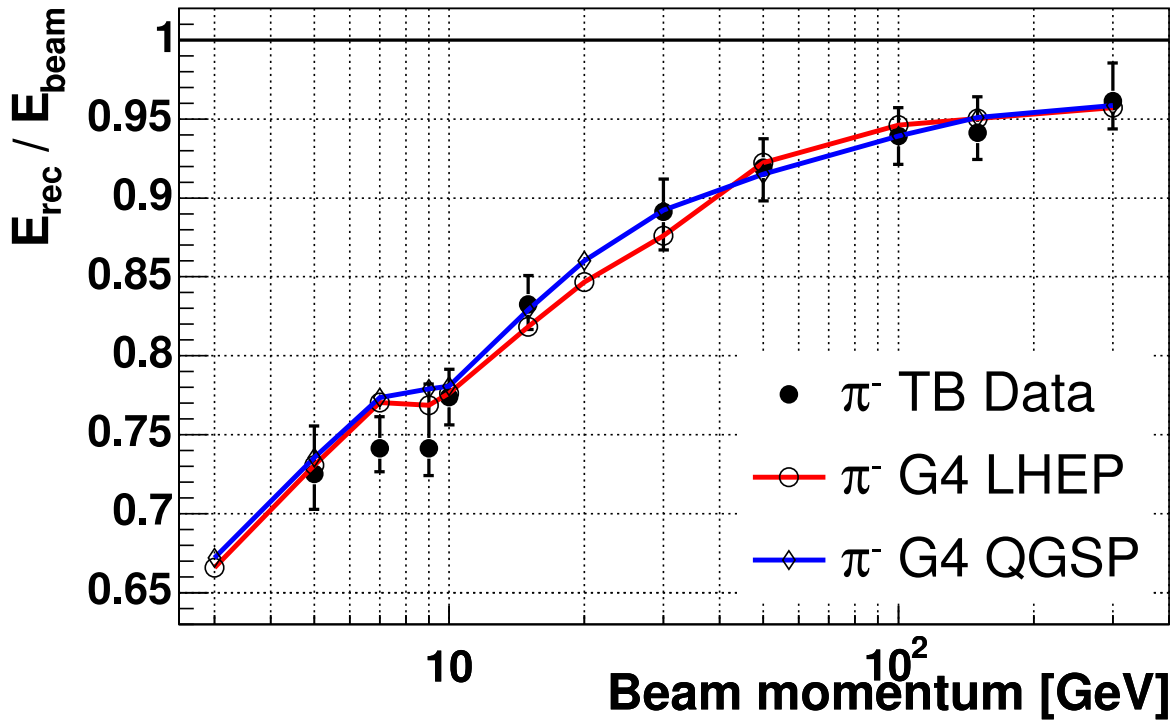}}
\caption
{CMS linearity plot with EMCAL and HCAL~\cite{piperov}
\label{cms1}}
\end{figure}

\begin{figure}[tbh!]
\centerline{\includegraphics[width=\textwidth]{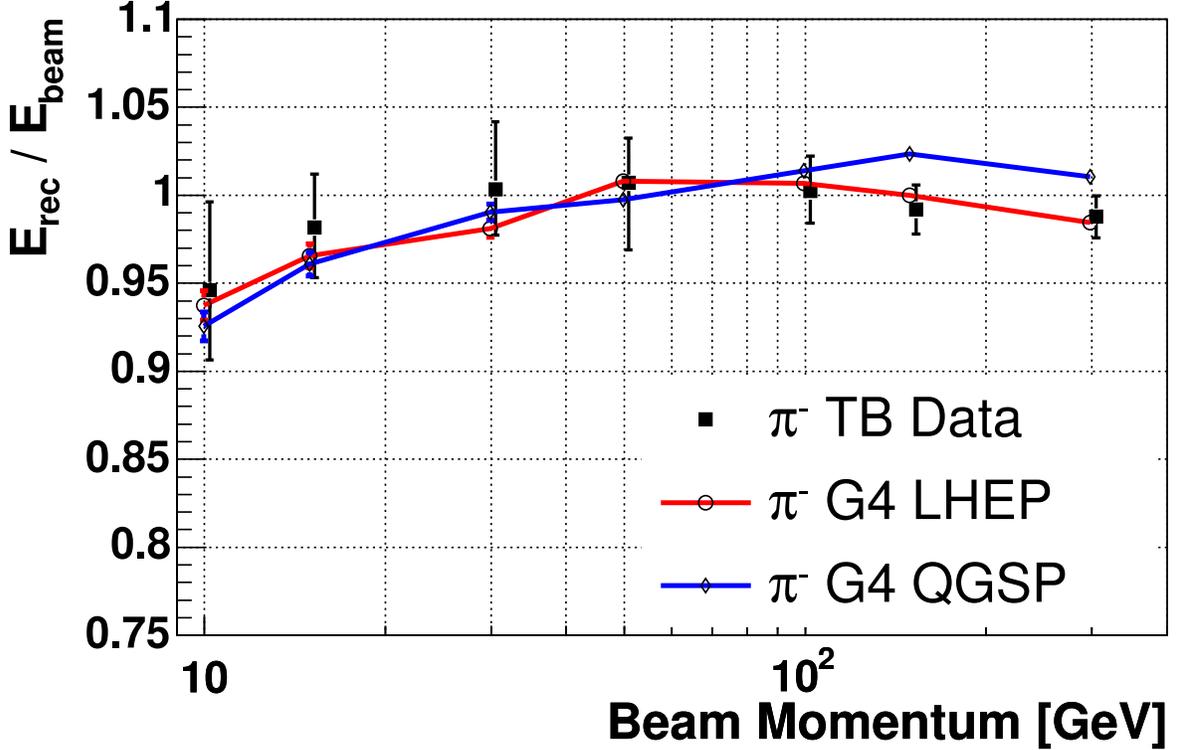}}
\caption
{CMS linearity plot with  HCAL only~\cite{piperov}
\label{cms2}
}
\end{figure}

The effect of calorimeter non-compensation can be seen in the CMS
detector, where two different technologies (crystals for EMCAL +
steel/scintillator for HCAL) produce significant lack of compensation
and non-linearities in the calorimeter hadronic
response. Figure~\ref{cms1} shows the significant non-linearity to
test beam pions with both EMCAL and HCAL and Figure~\ref{cms2} shows
the much better response of the hadronic calorimeter by
itself. Unfortunately while taking data, both EMCAL and HCAL come into
play for hadrons and non-linearities will effect jet and missing $E_T$
reconstruction. For the ILC, even if we succeed with the PFA
algorithm, a non-compensated calorimeter will make the response to the
neutral hadrons detected in the HCAL non-linear.

Calorimeter compensation can be achieved by two different techniques-
The first is by making the neutrons interact with hydrogenous material
that result in highly ionizing energy deposit from knock-on
protons. The fraction of hydrogenous material is carefully tuned to
achieve the right amount of boost in the hadronic response to
compensate for the energy lost in nuclear break-up. This approach was
used in the D\O\ calorimeter and the Zeus calorimeter both of which had
$e/\pi$ ratio close to unity. A single particle resolution of
$30\%/\sqrt(E)$ was achieved in the the Zeus calorimeter.  

The other approach for compensation is to obtain separate signals
from the electromagnetic energy and the hadronic energy 
(using fibers sensitive to Cerenkov and scintillator light) and weight
these two separately to optimize the resolution. This is the approach
taken by the DREAM calorimeter, the 4th concept of the ILC detector. 
We emphasize here that it is possible to have both a compensated
calorimeter and employ the PFA algorithm, if sufficient
care is taken to design and build such a calorimeter. Before we
propose a scheme that will serve both purposes, we point out the need
for longitudinal segmentation in a sampling calorimeter.
\subsection{The case for longitudinal segmentation} 
There exist two
different methods for calibrating sampling calorimeters. One is to
multiply the live energy in each layer by the inverse sampling
fraction yielding
\begin{equation}
E^{deposited} = \sum_{i=1}^{i=N_{layers}} \mu_i E_i^{live}
\end{equation}
This will give you the total energy on average but the resolution will
{\bf not} be optimal as has been shown in ~\cite{rajapaper}.  This is
a result of correlations between layers (longitudinal and transverse)
in a shower.

It can be shown that the best estimator of energy deposited in a
layer, that optimizes resolution~\cite{rajapaper} is given by
\begin{equation}
E^{deposited}_i = \sum_{j=1}^{j=N_{layers}} \lambda_{ij}E_j^{live}
\end{equation}
where the matrix $\lambda_{ij}$ (also referred to as the $\lambda$ tensor)
is in general non-diagonal.

\begin{figure}[tbh!]
\centerline{\includegraphics[width=\textwidth]{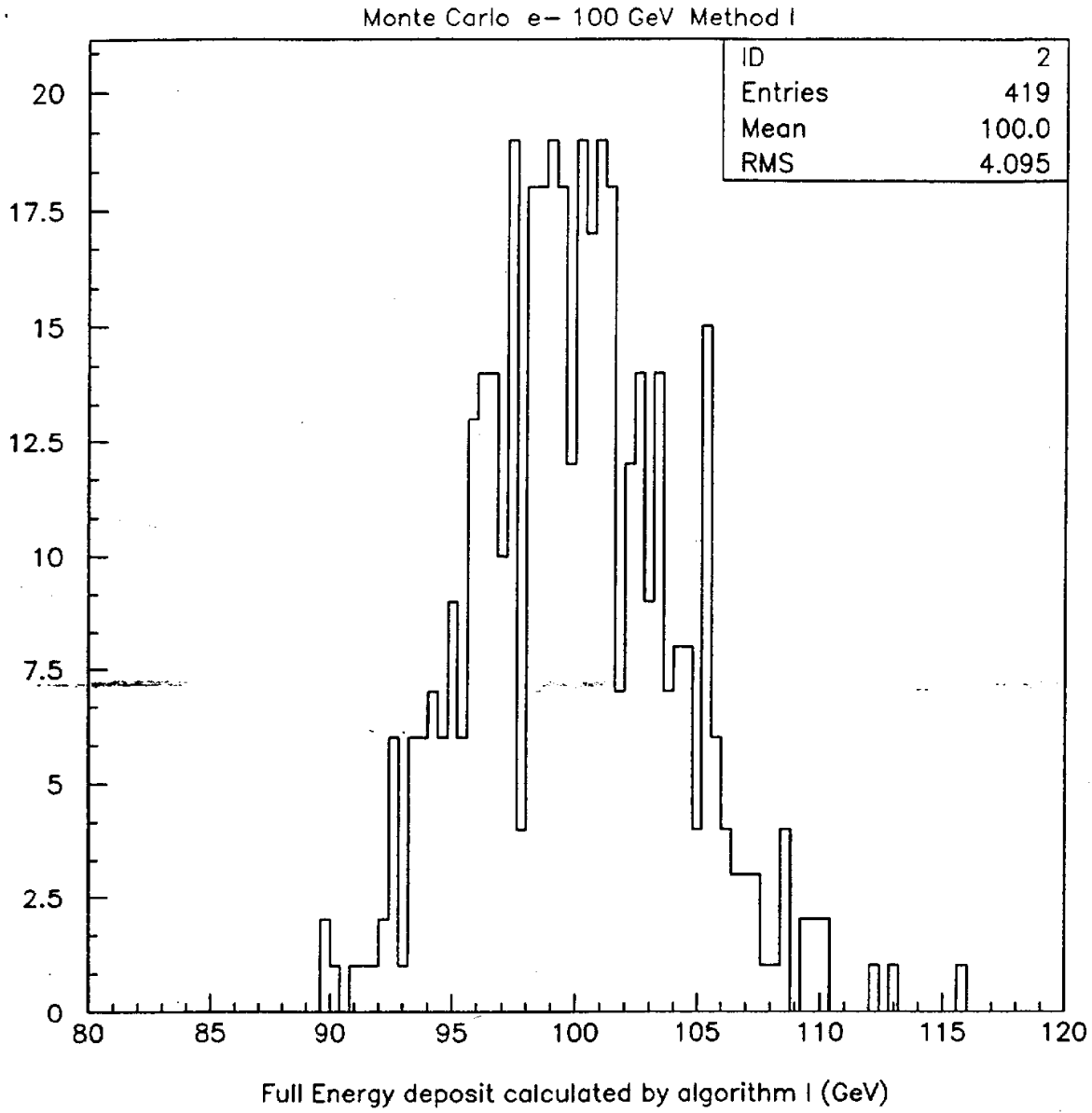}}
\caption
{Reconstructed energy of 100 GeV electrons using the inverse sampling
fraction method~\cite{rajapaper}
\label{raja1}
}
\end{figure}

\begin{figure}[tbh!]
\centerline{\includegraphics[width=\textwidth]{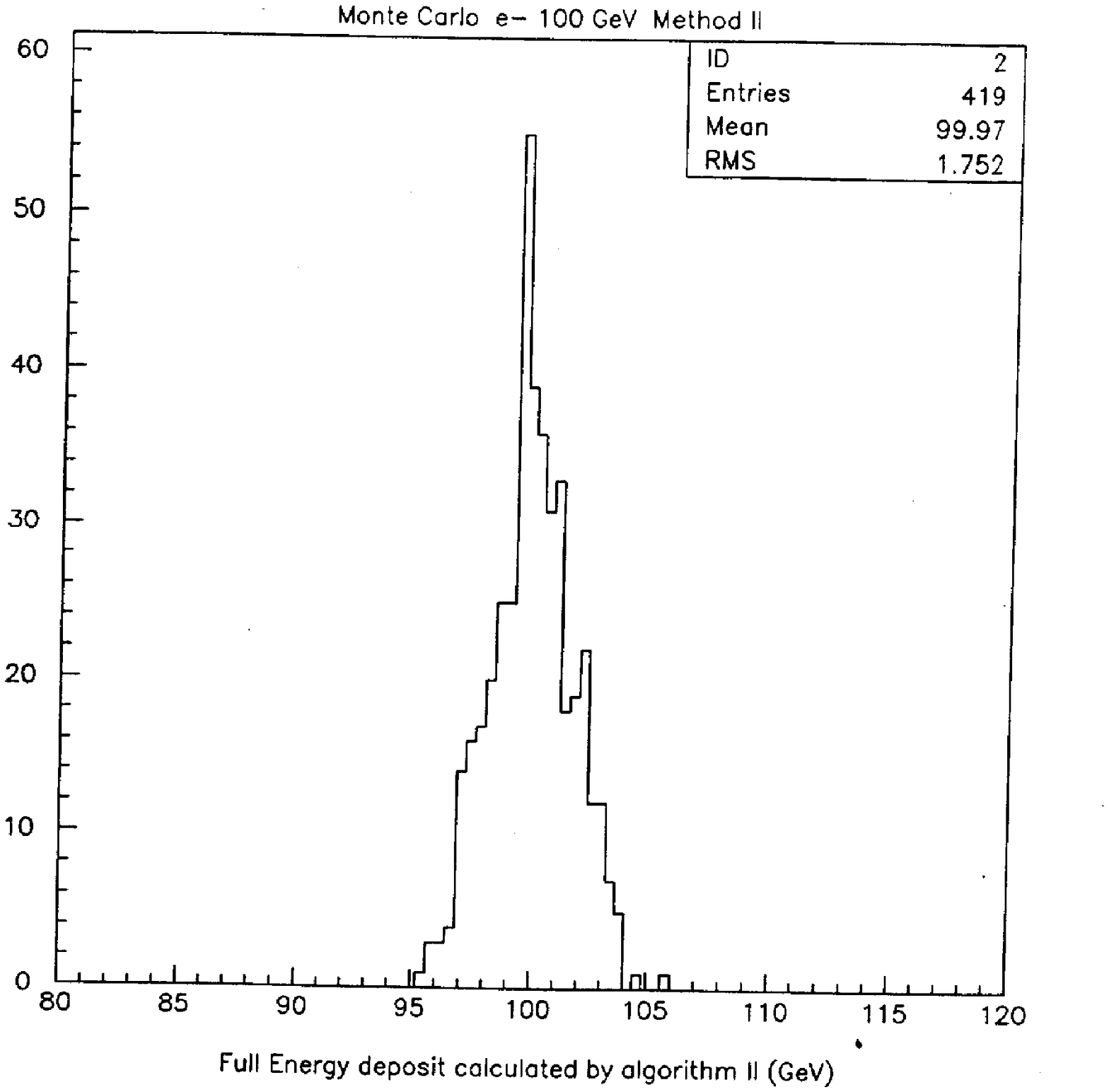}}
\caption
{Reconstructed energy of 100 GeV electrons using the weights
method~\cite{rajapaper}
\label{raja2}
}
\end{figure}

Then the best estimate of the deposited energy is
\begin{equation}
E_{deposited} = \sum_{j=1}^{j=N_{layers}} w_j E_j^{live}
\end{equation}
where the weights $w_i$ are in general not equal to the inverse
sampling fractions and are given by
\begin{equation}
w_j = \sum_{i=1}^{i=N_{layers}} \lambda_{ij}
\end{equation}
Figure~\ref{raja1} shows the reconstructed energy of 100~GeV electrons
using the inverse sampling fraction method and Figure~\ref{raja2}
shows the same data reconstructed using the weights
method~\cite{rajapaper}. It can be seen that the weights method gives
superior resolution. The formalism developed in~\cite{rajapaper} was
for EM showers, but applies equally to hadronic showers. In fact, it
can be shown that the inverse sampling fraction method and the weights
method become identical only in calorimeters where the sampling
fraction is unity and the $\lambda$ matrix is diagonal. In all other
cases, the weights method will give better resolution.

The $\lambda$ tensor can only be worked out if Monte Carlo information
is available. However, the weights $w_i$, which are used for
calibrating the calorimeter in practice can be derived from test beam
data using linear least squares minimization to optimize the
resolution. The weights, being the sum of the rows of the $\lambda$
tensor, are essentially non-local in nature and involve properties of
the entire calorimeter. They give better resolution by taking into
account shower correlations.
The DREAM~\cite{dream} calorimeter using fibers to read out EM and hadronic showers
will give superior resolution. It is however difficult to segment
longitudinally.

The case for longitudinal segmentation can be enhanced on pattern
recognition grounds as well.

\subsection{A possible way to achieve a highly 
segmented compensating calorimeter cheaply}
We now illustrate the possibilities of using a liquid argon TPC with
absorbers as a calorimeter. The whole physics case is predicated on
the premise that the event rate at the ILC is low enough that event
readout times of the order of a millisecond are tolerable. If this is
the case, the following scheme can be used to produce cheap, highly
segmented compensating calorimeters that will also serve the purposes
of the PFA algorithm.

\begin{figure}[tbh!]
\centerline{\includegraphics[width=\textwidth]{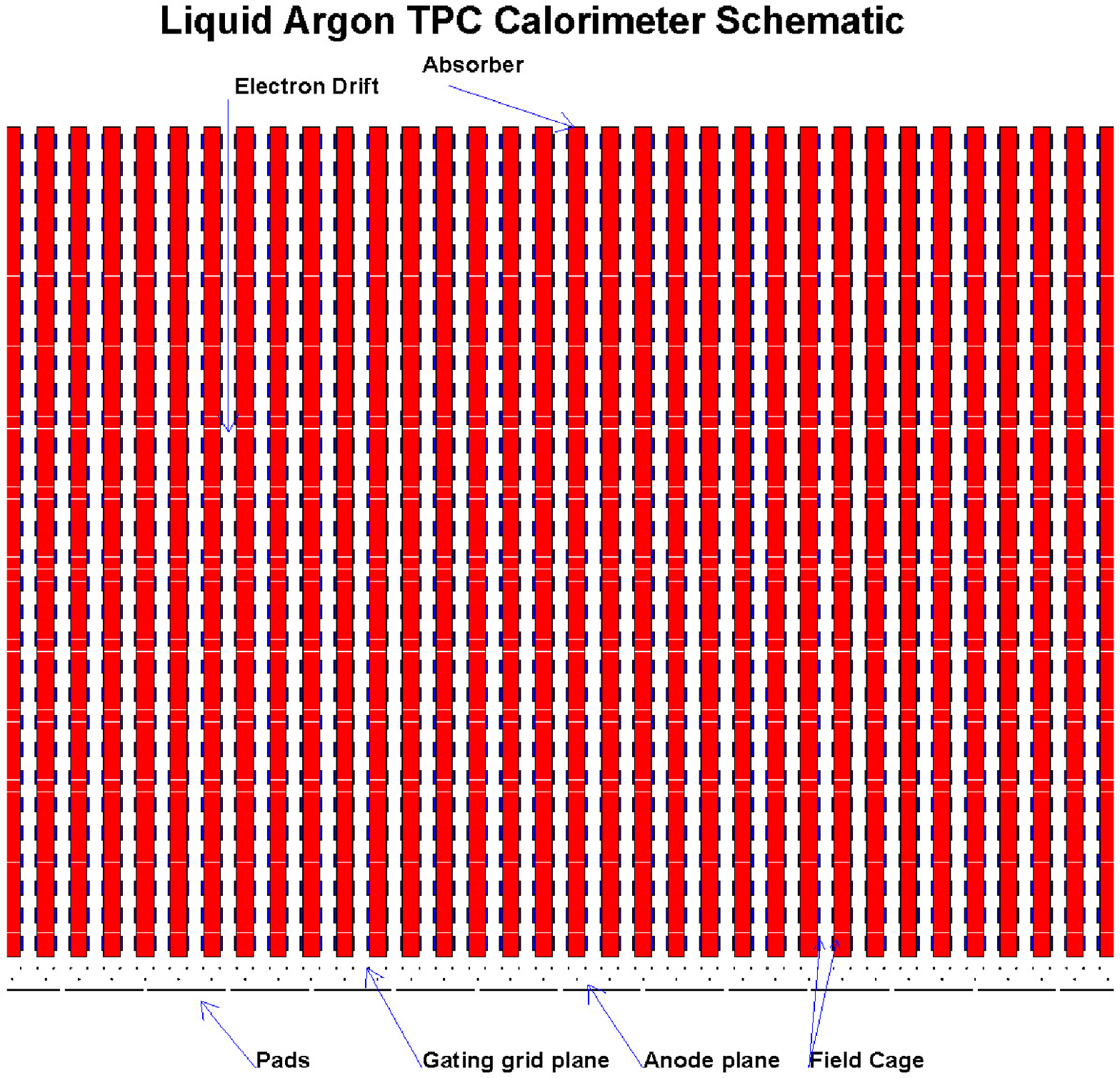}}
\caption
{Schematic of a liquid argon TPC calorimeter with absorber, field
cage, gating grid, anodes and pad shown. Beam enters left to right,
and electrons drift downwards. The absorber and liquid argon gaps are
comparable in size. The drift field is maintained in the gap by means
of a field cage in each gap. The longitudinal segmentation is governed
by the size of the pads and the transverse segmentation is provided by
the drift time measurement. Compensation is achieved by being able to
distinguish EM deposits and hadron deposits by the track density.
\label{ilccal}}
\end{figure}

Figure~\ref{ilccal} illustrates the scheme. Absorber plates of thickness
$\approx$ 3mm are immersed in liquid argon. The liquid argon gap is
also $\approx 3mm$.  It is probably best if the absorber material is
insulating. 
In each liquid argon gap, we construct a field cage using
copper strips engraved on a thin insulating sheet of material attached
to the absorber. This field cage will then support electric fields in
the argon of 500 V/cm. The drift velocity in argon is
1.5meters/millisecond. So if one is allowed a drift time of
0.5~millisecond, one can construct calorimeter modules of 75~cm long
in the drift direction.  The drifted charge is then measured by the
standard technique of a gating grid and anode wires and pads. The
longitudinal segmentation is governed by the size of the pads. In the
transverse direction, the time measurement will yield TPC like
resolution of the shower.

The diffusion of charge in liquid argon is $\approx$ 1.3~mm $\sigma$
after a drift time~\cite{kirkm} of 3~millisecond. This means that a liquid
argon gap size of 3~mm is sustainable for a drift time of 0.5~millisecond.

$E\times B$ effects in the drift will cause some distortion of the
track positions in the calorimeter as a function of drift time. In the
barrel region, the drift is along the solenoidal magnetic field, so
these effects are negligible. In the end-cap region, one can still
continue to drift along the field direction or better still one can
calibrate away the $E\times B$ effects.

The readout area has to be made as thin as possible, since it cannot
sustain absorbers and thus represents a crack. The effects on hermeticity 
of these
cracks can be ameliorated by staggering adjacent modules to avoid
pointing cracks.

The liquid argon cryostat is another area where hermeticity may be an
issue. This can be considerably ameliorated by using aluminum to
construct the cryostat and by using ``absorber gaps'' to read out the
shower immediately after the cryostat. Since the first piece of
material (ignoring the tracker) seen by the hadrons is the cryostat,
its effects on resolution are substantially reduced.

Compensation in such a device is attained by ``pattern recognizing''
(either online or offline) the electromagnetic showers by the local
density of tracks associated with them and separating them from
hadronic tracks this way. This then permits one to use two sets of
weights for EM and hadronic deposits leading to compensation. One can
fine tune the hydrogenous content in the calorimeter to tune out the
fluctuations in nuclear break up to obtain even better resolution.

Such a calorimeter, if realized, is cheap, highly segmented
longitudinally and transversely, compensating and suitable to be used
with the PFA algorithm.

We note that such a device would also serve as a far detector for a
neutrino factory since it can be made compact enough to sustain
magnetic fields affordably and the oscillation $\nu_e\rightarrow
\nu_\mu$ in the neutrino factory is detected by the appearance of wrong
sign muons. By reducing the absorber thickness and using appropriate 
absorber materials, one may even be able to use this device to 
search for rare decays of nuclei in the absorber.

The device proposed here will be comparatively inexpensive to build for a 
given segmentation than other technologies, since the time (drift ) 
dimension is as long as 75~cm and does  not need to be instrumented to 
obtain the segmentation.

The ILC detector concepts currently under consideration~\cite{ilcd}
consist of the GLD, LDC, and SiD detctors with the DREAM calorimeter
forming the center point of the 4th detector concept. The GLD concept
uses Tungsten (W)/Scintillator for its EM calorimeter, Pb/Scint for
the HCAL, and W/silicon and W/Diamond technologies for its forward
calorimeters. The LDC design has an EM calorimeter that uses
Silicon-Tungsten and a hadronic calorimeter made of iron/scintillator.
The SiD detector also employs Silicon-Tungsten for the EM calorimeter
and is doing R\&D on the hadronic calorimeter readout to choose between
GEM's. Scintillator, RPC's and silicon PMT's. All the detector groups
except the 4th concept employ the PFA algorithm to obtain good
resolution and the calorimetric designs are driven towards high
segmentation (and expense) based on the needs of the PFA algorithm.

\subsection{Test beam activity}

The calorimeter modules being constructed using various technologies
should be tested in charged particle beams that range in momentum from
$\approx$ $1~GeV/c$ to $100~GeV/c$. One needs to test the calorimeter 
response to both positive and negative beams with identified beam particles 
$\pi^\pm,K^\pm, p^\pm$).

Such beams are available at CERN
and at Fermilab M-Test area, where a new beamline has recently been
commissioned (shortened and upgraded, using experience gained in the 
MIPP beamline design) to provide lower energy beams.

Such an activity can help debug the electronics and test the design of
the modules by comparing the test beam response (longitudinal and
transverse) to hadronic shower simulation programs.  It is unlikely
that one hadronic shower simulation program out of the many
available can be used to fit all the data over all momentum and beam
species for all the technologies. One approach to take then would be
tune the parameters available in some of the models to a chosen
technology. But such a tuned model, if it works adequately will only
work for that technology and likely not for others. As mentioned previously,
improving shower simulation programs needs data at the hadron nucleus
level using a particle production experiment. The direct usage of events in 
simulation programs from
random access libraries will help reduce the model dependence considerably.

There is much less particle production data that have neutrons,
$K^0_L$ and anti-neutrons as beams compared to charged particle
production data. The neutral particle response in calorimeters is also
thus harder to predict well. The PFA algorithm proposes to measure the
charged particle energies using magnetic fields and to use the
calorimeter for neutral particle energy measurements (EM and
hadronic). In order to separate the neutral and charged hadronic deposits in
the calorimeter in a jet, it is thus necessary to understand the
transverse and longitudinal deposition characteristics of both the
charged and neutral particles.

It is here that the tagged neutral beam capabilities of an upgraded 
MIPP experiment can help the ILC effort.

\subsection{Tagged Neutral beams and ILC Detector R\&D}

\begin{figure}[htb!]
\centering
\includegraphics[width=\textwidth]{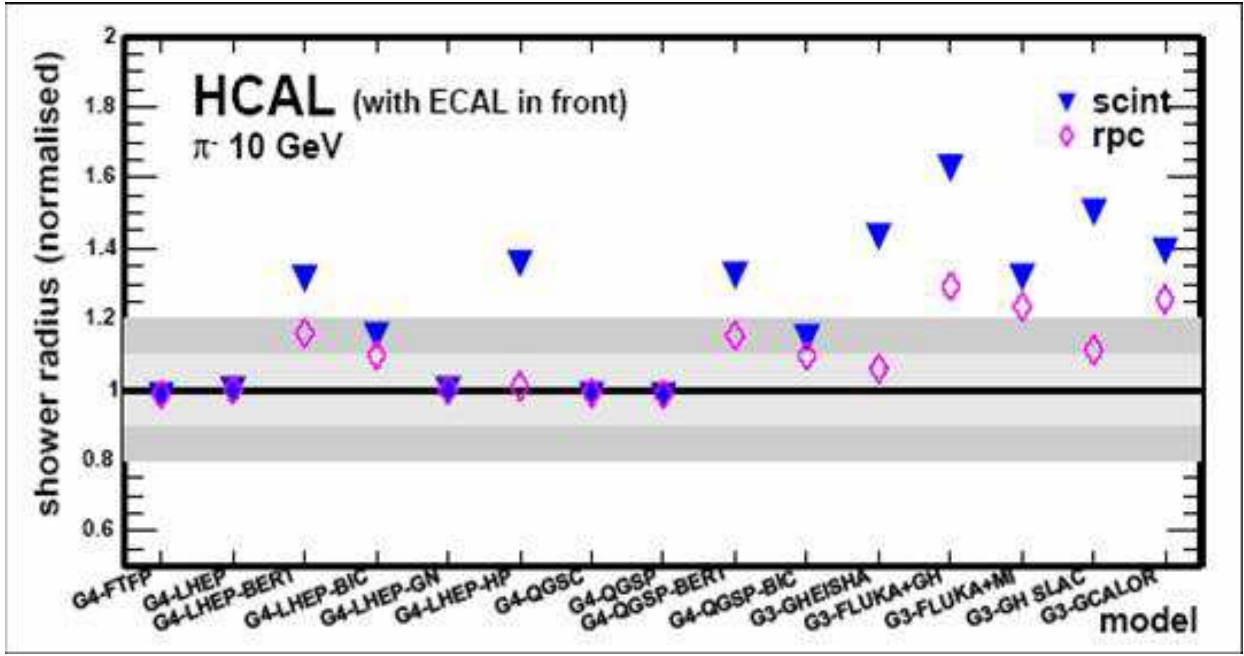}
 \caption{The width of a 10~GeV/c $\pi^-$ energy deposit in
 scintillator and RPC readout calorimeters as simulated by a host of
 simulation programs available in Geant4 (G4-) and Geant3 (G3-). The
 widths are normalized to the minimum width obtained.  }
\label{cwid}
\end{figure}

Three out of the four ILC detector concepts (SiD, LDC, GLD)~\cite{ilcd} are
optimized around the particle flow algorithm (PFA), which proposes to
measure the energy of jets in an event by using both the magnetic field
and the calorimeter. The charged particles are measured using the
excellent momentum resolution of the tracker  and the neutral
particles are measured using the calorimeter. The required fractional
energy resolution of a jet is $\sigma_E/E=0.3/\sqrt(E)$, E in GeV. This
hard-to-achieve performance is driven by the ILC design requirement to
be able to separate the processes $W\rightarrow jet+jet$ and
$Z\rightarrow jet+jet$. In order to measure the neutral particle
energy using the calorimeter, one needs to separate the charged
particle hits and the neutral particle hits in the calorimeter. This
dictates a highly segmented calorimeter. In order to test the design,
one needs to simulate the widths of the showers of both the charged
and neutral particles in the calorimeter.
Figure~\ref{cwid} shows the simulation of the width a 10~GeV $\pi^-$
particle entering two ILC calorimeters, one using RPC readout and the
other using scintillator readout~\cite{cwid-guy} for a variety of
hadronic shower simulators available in Geant4 and Geant3. The widths
are normalized to the narrowest width obtained. There is a variation
in the widths of $40\%$ in the simulations. This calls for a
data-based approach both for charged and neutral hadronic
responses. The upgraded MIPP spectrometer offers a unique opportunity
to measure the neutral particle response to three neutral species, the
neutron, the $K^0_L$ and the anti-neutron.

The basic idea is to use the diffractive reactions
\begin{eqnarray}
 pp\rightarrow n\pi^+p \\
 K^+p\rightarrow\bar{K^0}\pi^+p ; \bar{K^0}\rightarrow K^0_L\\
 K^-p\rightarrow{K^0}\pi^-p ; {K^0}\rightarrow K^0_L\\
 \bar{p}p\rightarrow \bar{n}\pi^-p 
\end{eqnarray}
where the beam of protons, $K^\pm$ or $\bar{p}$ fragments
diffractively to produce the neutral beam. The charged particles in
the reaction are measured in the MIPP spectrometer. The beam momentum
is known to $\approx$ 2\%. So the momentum of the tagged neutral
particle can be inferred by constrained fitting (3-C fit) to better
than $2\%$, event by event. The tagged neutral particle goes along the
beam direction and ends up in a test calorimeter placed in lieu of the
present MIPP calorimeter.

This technique demands that the target is a proton and will only work
on a liquid hydrogen cryogenic target (that MIPP possesses). The
plastic ball recoil detector in the MIPP Upgrade
will act as an additional veto against
neutral target fragments such as slow $\pi^{0'}s$.

\begin{figure}[tbh]
\begin{center}
\includegraphics[width=\textwidth]{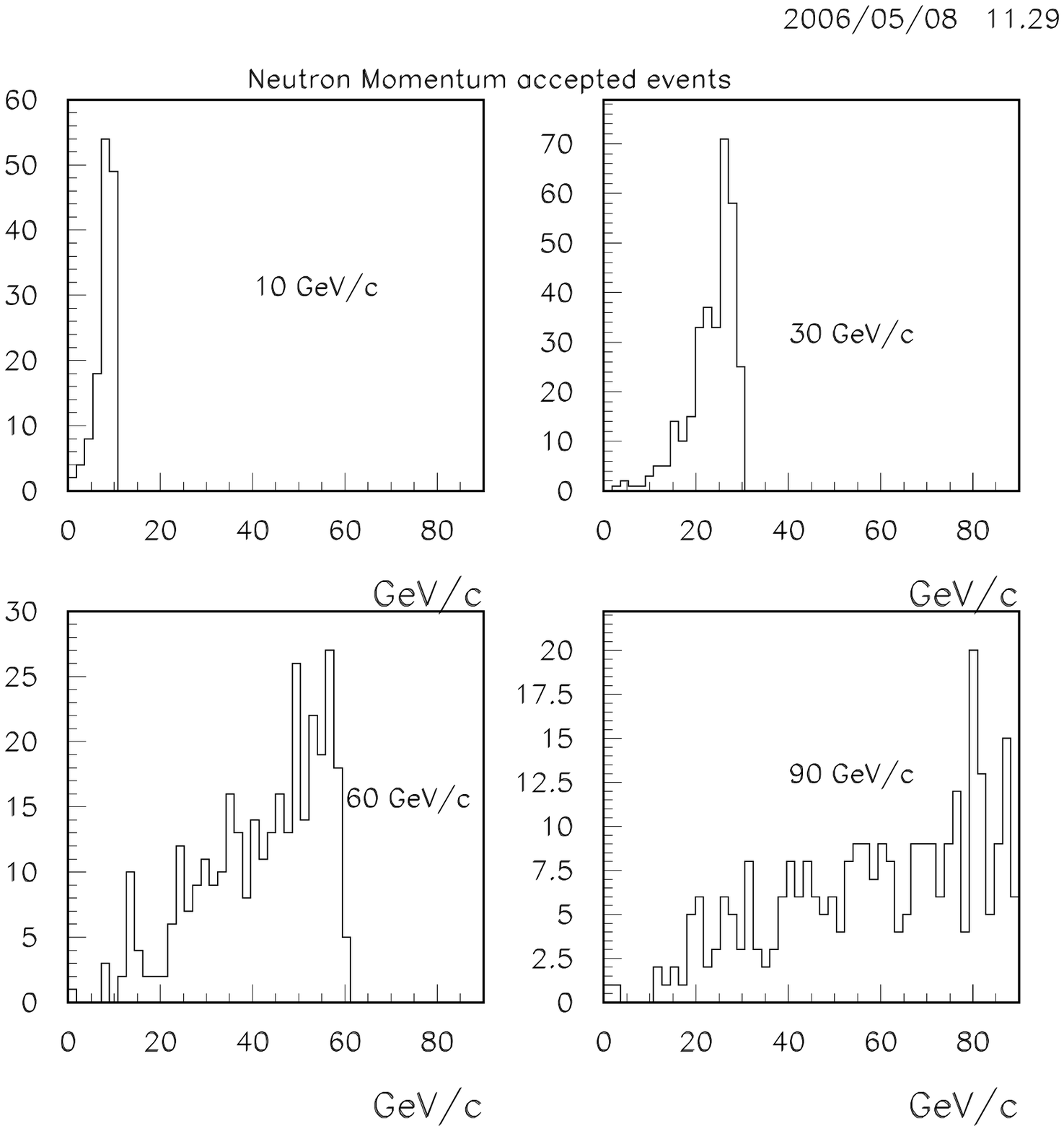}
\caption{Momentum spectrum  of accepted neutrons for incident proton momenta of
10~GeV/c, 30~GeV/c, 60~GeV/c and 90~GeV/c 
for the process $pp\rightarrow pn\pi^+$.~\label{pneut}}
\end{center}
\end{figure}

The momentum spectrum of the neutral beam is controllable by changing the beam
momentum. The method is outlined in detail in  MIPP note 130~\cite{raja130}.
The diffractive processes are simulated using the program DPMJET and the event rates estimated for a calorimeter placed in the MIPP calorimeter position. 
With the MIPP upgrade, it should be typically possible to obtain 
$~\approx$~50,000 tagged neutrons, 
$\approx$~9,000 tagged $K^0_L$, and 
$\approx$~11,000 tagged $\bar n$  per day in the calorimeter with the
 beam momentum set to 20~GeV/c. Table~\ref{tagt} shows the expected 
number of events /day as a function of beam momentum and beam species.
\begin{table}
\caption{Expected number of tagged neutrons, $K^0_L$, and anti-neutrons per day with an upgraded MIPP spectrometer.\label{tagt}}
\begin{tabular}{|c|c|c|c|c|}
\hline
Beam Momentum & Proton beam & $K^+$ beam & $K^-$ beam & ${\bar p}$ beam \\
\hline
GeV/c         & n/day    & $K^0_L$/day   & $K^0_L$/day & ${\bar n}$/day \\
\hline
10 &  20532 & 4400 & 4425 & 6650\\
\hline
20  & 52581 & 9000 & 9400 & 11450\\
\hline
30 &  66511 & 12375 & 14175 & 13500\\
\hline
60 &  47069 & 15750 & 14125 & 13550\\
\hline 
90 &  37600 &- &- & \\
\hline
\end{tabular}
\end{table}
Figure~\ref{pneut} shows the momentum spectrum of tagged neutrons
accepted in the calorimeter as a function of the beam momentum. Other
similar plots are available in MIPP note 130~\cite{raja130}.
The acceptance of the calorimeter is assumed to be a circle of radius
75~cm about the beam axis, for the purposes of this calculation. The
event is accepted only if a recoil proton of energy greater than
200~MeV is produced that is detected in either the plastic ball or the
TPC.  The rates quoted here are for an upgraded MIPP spectrometer
operating at 3kHz DAQ rate, with a Main Injector spill of 4~sec
duration delivered every 2 minutes and a beam downtime of
$42\%$. These beam rates produce 12,000 interactions per spill (5
million interactions/day) which are recorded and stored on mass
storage device. These events can be used for physics. The ILC
calorimeter can be triggered using a ``neutral energy deposited ``
trigger and the calorimeter hits are read out only for those
events. These events (maximum rates for inclusive neutrons are ~3000
per spill) can be filtered and stored on a separate file in real
time. The tagged neutral events can then be extracted using
3-Constraint kinematic fitting. The fitting will lead to an energy
resolution event by event for each neutral particle entering the
calorimeter of $\approx$~2\%.

\subsection{Tagged $\pi^0$'s}

The technique outlined here can also be used to produced tagged beam
of $\pi^0$'s, which can be used to study the $\pi^0$ mass
reconstruction capabilities of the ILC calorimeter. For this purpose, we
employ beams of charged pions and employ the diffractive channels
\begin{eqnarray}
\pi^+p\rightarrow \pi^0\pi^+ p \\
\pi^-p\rightarrow \pi^0\pi^- p \\
\end{eqnarray}
The outgoing $\pi^0$ momentum can be solved using a 1-C fit, since we
do not know the impact point of the $\pi^0$. The mean momentum of the
$\pi^0$ can be varied by changing the incoming beam momentum and in the
cases where the $\pi^0$ decays in a fashion such that both photons end
up in the calorimeter, one can solve for the energy of the two photons
given their impact point in the calorimeter. One can also compare
the calorimeter reconstructed $\pi^0$ mass to PDG values.
\subsection{Simulation of jets in a test beam}

So far we have only considered the problem of measuring the single
particle response of calorimeters. An underlying assumption is that we
can superpose single particle responses of particles to build the
response to jets of particles. This assumes linearity of response and
ignores problems such as coherent electronics noise etc.

Jets at the ILC consist of particles resulting in the fragmentation of
all quark types (u,d,c,s,t,b). In the test beam, we only have access
to particles made up of $u,d,s $ quarks. Thus it is 
not feasible  to reproduce the particle content and multiplicities of
jets that will be encountered at the ILC in a test beam. One can
however measure single particle response of the calorimeters to the
six charged beam species $\pi^\pm,K^\pm,p^\pm$ as well as the tagged
neutral beams $n,{\bar n}, K^0_L$ and $\pi^0$'s. One can then use these
single particle data to simulate the behavior of ILC jets assuming
linear supersposition holds.  Let us note that such studies will only
describe the ILC jets in regions of phase space (dead material, angles
of incidence) similar to the test beam calorimeter and will not in
general simulate the effect of ILC tracking detector magnetic
fields. However, this gives an idea of the behavior of jets in a 
full ILC detector.

Linear superposition is not guaranteed  
in calorimeters. Electronic cross-coupling
(coherent noise) and saturation effects will in general work
against such an assumption. So it may be desirable to study the 
response of the calorimeter prototype 
to multiple tracks hitting it simultaneously.
This can be done by  using the 5~million events/day
written out at the tagged neutral beam. 
Events in which the beam particle dissociates diffractively contain
forward going charged and neutral
particles. The tagged neutral beam events form a subset of a much larger set
of events in which forward going multiple particles exist. 
The MIPP spectrometer is so configured that the downstream
magnet (ROSIE) bends the particles back in opposite direction to the
Jolly Green Giant magnet in which the TPC sits. This results in a
large acceptance of particles above $\approx$ 7~GeV/c at the RICH
detector and consequently the ILC calorimeter which will sit
downstream of the detector. All these charged particles will be
identified by the MIPP spectrometer and multiplicities of 4-5 at the
RICH counter are quite common. So the 5~million events written out/day  can
be a rich source for studying multi-particle effects in the
calorimeter. The momentum content of these particles can be varied by
changing the MIPP primary beam momentum.

\section{Conclusions}
We have outlined a two pronged approach to understanding the problem
of understanding hadronic showers in calorimeters for the ILC. It
entails obtaining new data to improve the shower simulation programs
while simultaneously measuring the response to charged and tagged
neutral beams to calorimeters.

We have outlined a scheme to construct a compensating highly segmented
calorimeter cheaply that can be made compensating and also serve the
interests of the PFA algorithm. 

The program is ideal for training graduate students in the techniques
employed in HEP. The impact of such a program will be beyond the ILC
itself and will help cosmic ray energy systematics, atmospheric
neutrino experiments and the ongoing worldwide neutrino oscillation
program.
\section{Acknowledgements} 
The author wishes to thank Robert
Tschirhart for asking that such a paper as this be put together, 
Marcel Demarteau for general
encouragement, Nikolai Mokhov and Sergei Striganov for providing the
Monte Carlo validation plots and Stephen Pordes for conversations
involving liquid argon TPC's. This work was supported by the
U.S. Department of Energy.

\end{document}